\documentclass[lettersize,journal]{IEEEtran}

\usepackage[numbers,sort&compress]{natbib}
\usepackage{multirow}
\usepackage{threeparttable}
\usepackage{amsmath,amsfonts,amssymb}
\usepackage{graphicx}
\usepackage{subfig}
\usepackage{algorithmicx}
\usepackage[ruled,linesnumbered]{algorithm2e}
\usepackage{algpseudocode}
\usepackage{textcomp}
\usepackage{makecell}
\usepackage{multirow,multicol}
\usepackage[colorlinks]{hyperref}
\hypersetup{citecolor=blue}
\usepackage[table]{xcolor}
\usepackage{balance}
\usepackage{threeparttable}
\usepackage{amsthm}
\usepackage{color}

\begin{document}
	
	\title{Integrated Sensing and Communications for Unsourced Random Access: A Spectrum \\Sharing Compressive Sensing Approach}
	\author{Zhentian Zhang, Jian Dang, Kai-Kit Wong, Zaichen Zhang and Christos Masouros
		% <-this % stops a space
		\thanks{ }% <-this % stops a space
		\thanks{Zhentian Zhang, Jian Dang, Zaichen Zhang are with the National Mobile Communications Research Laboratory, Frontiers Science Center for Mobile Information Communication and Security, Southeast University, Nanjing, 210096, China. J. Dang and Zaichen Zhang are also with the Purple Mountain Laboratories, Nanjing 211111, China. (e-mails: \{zhangzhentian, dangjian, zczhang\}@seu.edu.cn).}% <-this % stops a space
		\thanks{Kai-Kit Wong and  Christos Masouros are with the Department of Electronic and Electrical Engineering, University College London, Torrington Place, WC1E 7JE, United Kingdom and Kai-Kit Wong is also with the Yonsei Frontier Lab., Yonsei University, 03722 Korea. (e-mails: \{kai-kit.wong, c.masouros\} @ucl.ac.uk).}}
%		\thanks{Corresponding authors: Zaichen Zhang (zczhang@seu.edu.cn); J. Dang (dangjian@seu.edu.cn)}}% <-this % stops a space
%	
%	\author{Zhentian Zhang
%		        % <-this % stops a space
%		\thanks{This paper was produced by the IEEE Publication Technology Group. They are in Piscataway, NJ.}% <-this % stops a space
%		\thanks{Manuscript received April 19, 2021; revised August 16, 2021.}}
	
	 \pagestyle{empty}
	\maketitle
	  \thispagestyle{empty}
	\begin{abstract}
		This paper addresses the unsourced/uncoordinated random access problem in an integrated sensing and communications (ISAC) system, with a focus on uplink multiple access code design. Recent theoretical advancements highlight that an ISAC system will be overwhelmed by the increasing number of active devices, driven by the growth of massive machine-type communication (mMTC). To meet the demands of future mMTC network, fundamental solutions are required that ensure robust capacity while maintaining favorable energy and spectral efficiency. One promising approach to support emerging massive connectivity is the development of systems based on the unsourced ISAC (UNISAC) framework. This paper proposes a spectrum-sharing compressive sensing-based UNISAC (SSCS-UNISAC) and offers insights into the practical design of UNISAC multiple access codes. In this framework, both communication signals (data transmission) and sensing signals (e.g., radar echoes) overlap within finite channel uses and are transmitted via the proposed UNISAC protocol. The proposed decoder exhibits robust performance, providing 20-30 dB capacity gains compared to conventional protocols such as TDMA and ALOHA. Numerical results validate the promising performance of the proposed scheme.
	\end{abstract}
	\section{Introduction}
	\subsubsection{Background and Related Work}
	Massive machine-type communication (mMTC) is deemed as one of the most crucial application scenarios for the future communication networks. Unfortunately, the coordinated multiple access protocols \cite{URA_China_Comm} become invalid dealing with dense number of cheap devices. From information theory perspective \cite{WYP_Survey}, the users' average channel capacity will converge to zero as the number of devices approaches infinity, even if other system metrics, such as sum-rate, are satisfactorily optimized.
	
	To tackle the aforementioned issue, an uncoordinated multiple access technique called unsourced random access (URA) is posed in \cite{Polyanskiy} and has the potentiality to support unbounded number of users with an optimized concatenated multiple access code design. Recent advances on information theory aspect can be referred in
%	 \cite{Bound_with_random_activity,fading_MAC,Bound_MIMO2,Bound_with_Quasi_Channel}, 
	 	 \cite{Bound_with_random_activity,Bound_with_Quasi_Channel}, where achievable bounds for different channel models are established. While many state-of-the-arts solely focus on communication task, integrated sensing and communication (ISAC) has become an inevitable trend for the future networks \cite{ISAC1}. Recent work \cite{ISAC-URA} demonstrates that an ISAC system will be overwhelmed by the surging number of users, unveiling the challenges of unsourced ISAC (UNISAC) and providing theoretical benchmarks for the practical design of UNISAC multiple access codes. The novel UNISAC architecture demonstrates significant capacity gains compared to traditional multiple access protocols, such as time division multiple access (TDMA), ALOHA, and treating interference as noise (TIN). UNISAC offers favorable features in terms of energy and spectral efficiency, while conventional protocols either fail to support the increasing activity density or require intolerable energy consumption. Yet, the practical design of UNISAC remains an open topic.
 \subsubsection{Motivations}
 Currently, communications-radar spectrum sharing (CRSS) \cite{CRSS1,CRSS2} is considered one of the most promising methods for enabling efficient spectrum utilization and for designing novel systems that benefit from the cooperation between radar and communications. In CRSS, dual-functional waveforms support sensing and detection while simultaneously carrying information. The concept of CRSS aligns closely with the UNISAC approach \cite{ISAC-URA}, where both sensing and communication signals overlap over a fixed duration of channel use. This resource-sharing approach, which provides access to all resources across potential users, maximizes the degrees of freedom for future network architectures.
 
The information for sensing and communication is projected onto distinct common codebooks, and the receiver must decouple the overlapping signals to simultaneously decode the information carried by the transmitted codewords and detect sensing targets. Notably, only a small portion of the codewords is activated at any given duration, introducing a sparsity feature. This sparsity in the code domain aligns with the design principle of compressive sensing (CS), enabling robust solutions for both linear and nonlinear regression models.
\subsubsection{Contributions}
Although the theory of UNISAC has provided system metric benchmarks, the design of a practical scheme remains an open challenge. Compressive sensing (CS)-based schemes \cite{CS_URA1,CS_URA2} demonstrate robust anti-disturbance capabilities and effective collision-error resolution. Treating the overlapped signals as a sparse recovery problem naturally aligns with the structure of a CS-based receiver.
 
Inspired by the aforementioned approach, we propose a spectrum-sharing compressive sensing method for UNISAC (SSCS-UNISAC), leveraging the sparsity of the signal components. Specifically, the non-linear projection of binary messages onto spreading sequences or pilots significantly reduces multi-user interference and provides practical solutions for multi-user detection. The sensing and communication users are effectively separated in both the code and power domains through non-orthogonal transmission. Furthermore, the complexity of the main decoding procedures is analyzed. Extensive numerical results validate the potential of the proposed scheme and compare it with the achievable limits of both UNISAC and conventional protocols.
\subsubsection{Content Structures}
In Sec.~\ref{sec.2}, the system description and encoder design for the proposed SSCS-UNISAC scheme are explained. In Sec.~\ref{sec.3}, the proposed decoder design is elaborated. In Sec.~\ref{sec.4}, extensive numerical results are presented to demonstrate the superiority of the proposed scheme compared to conventional protocols. Finally, conclusions are drawn in Sec.~\ref{sec.5}.
 \begin{figure}[htp]
 	\centering
 	\includegraphics[width=3.5in]{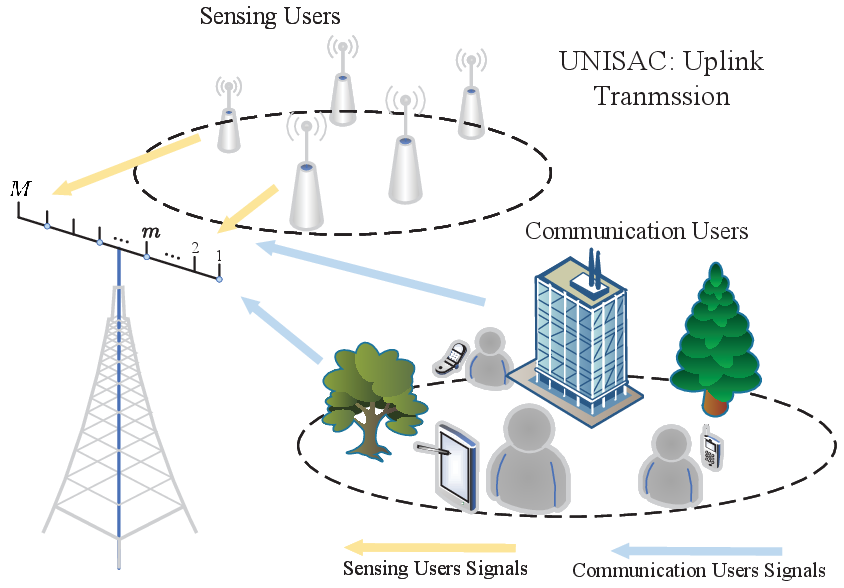}
 	\caption{Illustration of uplink transmission in UNISAC system.}
 	\label{system_model}
 \end{figure}
\section{SSCS-UNISAC System Model}\label{sec.2}
\subsection{System Descriptions}
This work considers uplink transmission with active sensing in a manner of UNISAC illustrated in Fig. \ref{system_model}. Let set $\mathcal{A}_c$ denote the binary message list of communication users (CUs) and set $\mathcal{A}_s$ contains all the angles of arrival (AOAs) of sensing users (SUs), i.e., $\theta_i \in \mathcal{A}_s$. Assuming there are $|\mathcal{A}_c|$ CUs and $|\mathcal{A}_s|$ SUs with single antenna served by an $M$-antenna receiver, each CU intends to send $B$ bits of information with $L$ channel uses and each SU selects a pilot from sensing common codebook via $B_s$ bits. Omitting asynchronous errors, signals of CUs and SUs are overlapped and received by the receiver. The system follows the power constraint of energy-per-user $E/N_0$ aligning with work \cite{ISAC-URA}:
\begin{equation}
	\label{eq:1}
	\frac{E}{N_0}=\frac{P}{\left(|\mathcal{A}_c|+|\mathcal{A}_s|\right)\sigma^2},
\end{equation}
where scalar $\sigma^2$ denotes the variance of the background additive white Gaussian noise (AWGN) and scalar $P$ is the total power of all users. We use $\beta\in [0:1]$ to denote the ratio of power allocated to sensing users, i.e., single CU follows the power constraint of $\frac{(1-\beta)P}{|\mathcal{A}_c|}$ and each SU of $\frac{\beta P}{|\mathcal{A}_s|}$.

The system metrics are denoted by the per-user probability error (PUPE) and AOA mean square error (AOAMSE):
\begin{equation}
	\label{eq:2}
	\begin{aligned}
	\mathrm{PUPE} &= \frac{|\mathcal{L}_{\mathrm{c,md}}|+|\mathcal{L}_{\mathrm{s,md}}|}{|\mathcal{A}_c|+|\mathcal{A}_s|},\\
	\mathrm{AOAMSE} &= \frac{1}{\tilde{\mathcal{L}}_{s,d}}\sum_{i\in \tilde{\mathcal{L}}_{s,d}}\mathbb{E}\left\{|\cos\theta_i-\cos\tilde{\theta_i}|^2\right\},
	\end{aligned}
\end{equation}
where sets $\mathcal{L}_{\mathrm{c,md}}$ and $\mathcal{L}_{\mathrm{s,md}}$ denote the missed detection of CUs and SUs, set $\tilde{\mathcal{L}}_{s,d}$ denotes the detected AOA list of SUs and angles $\tilde{\theta}_i\in \tilde{\mathcal{L}}_{s,d}$. The UNISAC receiver aims to reach a set of goals (PUPE and AOAMSE) with energy-per-user as low as possible. Meanwhile, due to the finite common codebook size, the codeword collision-derived error is an inherent issue for UNISAC when the information restoration hinges on codeword selection. If collision brings about decoding error, it will automatically be reflected by PUPE. While part of the collision error can be mitigated using channel coding\cite{TVT1}, the fundamental solution for practical scheme design is to enlarge the codebook size, making the collision error negligible.

We use $\mathbf{h}_k\in \mathbb{C}^{1\times M}, k\in \left[ 1:|\mathcal{A}_c|+|\mathcal{A}_s| \right]$ to denote the channel coefficients between users and the receiver. For CUs, the channel coefficient element distributes as $\mathcal{CN}(0,1)$. The Raleigh fading model can be commonly seen in environment with significant obstructions leading to multiple reflections and scattering, such as urban environment. For SUs, they might be static sensors deployed close to the receiver, typically experiencing a dominant line-of-sight (LOS) component. This setup aligns well with the uniform linear array (ULA) channel model. For ULA with half-wavelength rule\cite{ISAC-URA}, i.e., the gap distance equals half wavelength $d=\frac{\lambda}{2}$, the channel vector can be expressed as:
\begin{equation}
	\label{eq:3}
	\left[1,e^{-j\pi\cos\theta_k},\ldots,e^{-j\pi(M-1)\cos\theta_k}\right].
\end{equation}
 \begin{figure*}[htp]
	\centering
	\includegraphics[width=7in]{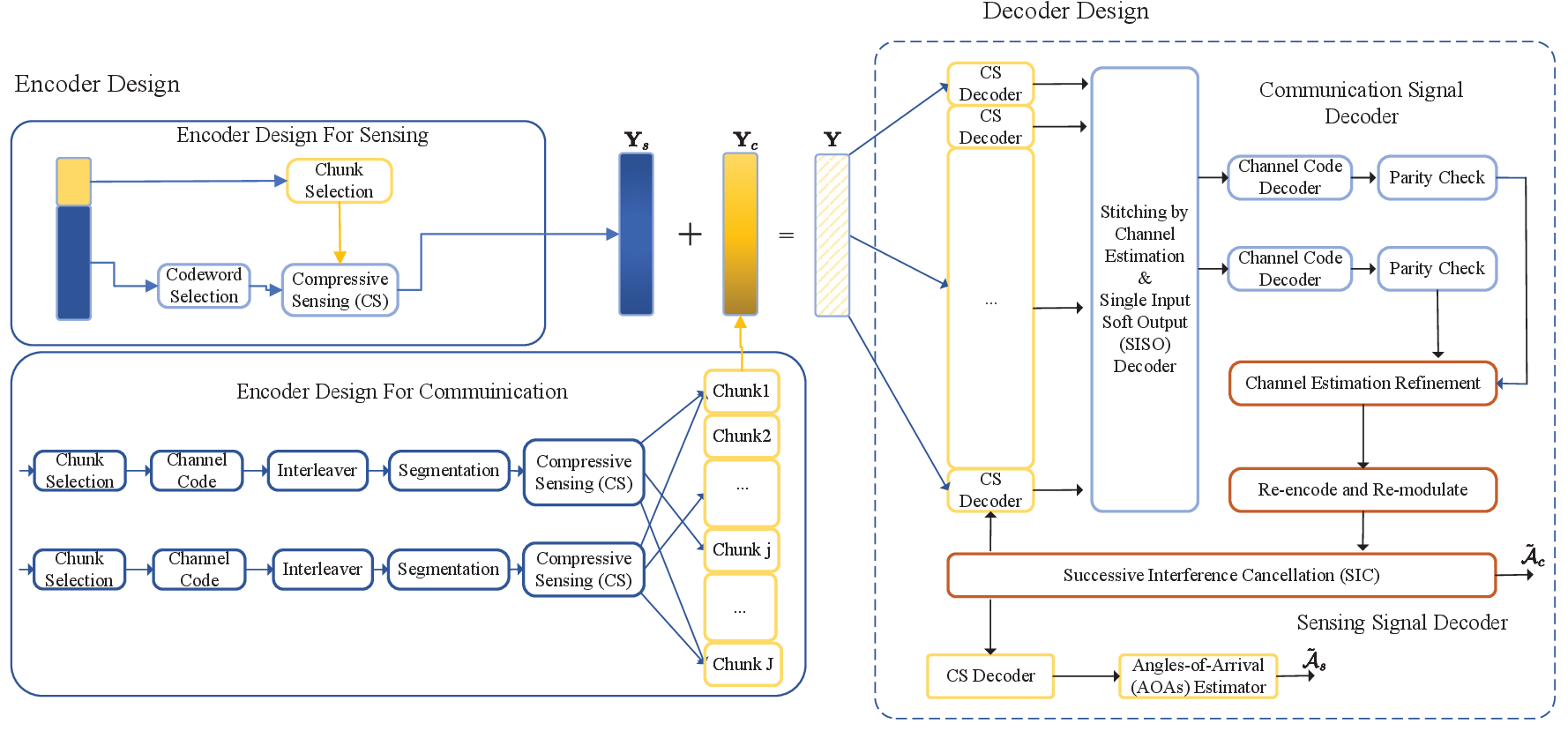}
	\caption{Illustration of the proposed UNISAC encoder and decoder designs. Notably, the encoder design for sensing does not imply that the SUs are coordinated or intend to transmit data. The sensing signals can be equivalently treated as simultaneous radar echoes from multiple targets.}
	\label{Encoder_Decoder}
\end{figure*}
\subsection{Encoder Design}
The overall transmission is uniformly divided into $J_c$ chunks/sub-frames with $\tilde{L}=\frac{L}{J_c}$ channel uses and every user selects a chunk to transmit the signal. The encoder design will explain how the user chooses chunks and encodes their binary messages. The proposed UNISAC encoder and decoder designs are illustrated in Fig. \ref{Encoder_Decoder}.
\subsubsection{SUs' Encoder Design}
Let $\mathbf{u}_{k,s} \in \{0,1\}^{B_s}$ denote the $k$-th SU's binary bits divided into $B_{\mathrm{chunk}}$ and $B_{p,s}$, i.e., $B_s=B_{\mathrm{chunk}}+B_{p,s}$. Each SU selects a chunk for transmission by the decimal value of $B_{\mathrm{chunk}}=\log_2J_c$ bits. Then, by the decimal value of $B_{p,s}$ bits, SUs select a pilot codeword from sensing codebook $\mathbf{A}_s=\left[\mathbf{a}_{s,1},\mathbf{a}_{s,2},\ldots,\mathbf{a}_{s,2^{B_{p,s}}}\right] \in \mathbb{C}^{\tilde{L}\times 2^{B_{p,s}}}$ where codeword follows the power constraint of $\|\mathbf{a}_{s,i}\|_2^2=\frac{\beta P}{|\mathcal{A}_s|},i\in [1:2^{B_{s,p}}]$. Thereby, the signal component of SUs at given transmission chunks can be written as $\mathbf{Y}_s = \sum_{k=1}^{\bar{K}_s}\mathbf{x}_{s,k}\mathbf{h}_{s,k}$, where vector $\mathbf{x}_{s,k}$ denotes the signal a SU transmit and scalar $\bar{K}_s$ denotes the number of SUs at given chunk.
\subsubsection{CUs' Encoder Design}
Let $\mathbf{u}_{k,c}\in \{0,1\}^{B}$ denote the $k$-th CU's binary message bits. CUs' $B$ bits are split into two portion of $B_{\mathrm{chunk}}$ and $B_c$ bits, i.e., $B=B_{\mathrm{chunk}}+B_c$. The transmission chunk is selected via $B_{\mathrm{chunk}}$ bits. The rest $B_c$ bits are then channel code encoded into $E$ bits, interleaved and then uniformly divided into $J$ segments with $B_p$ bits each, i.e., $B_p=\frac{E}{J}$. Each binary segment will be used to select a pilot codework from communication common codebook $\mathbf{A}_c\in \left[\mathbf{a}_{c,1},\mathbf{a}_{c,2},\ldots, \mathbf{a}_{c,2^{B_p}}\right] \in \mathbb{C}^{L_p \times 2^{B_p}}$ where scalar $L_p=\frac{\tilde{L}}{J}$ and each codeword follows the power constraint of $\|\mathbf{a}_{c,i}\|_2^2=\frac{(1-\beta) P}{J|\mathcal{A}_c|},i\in [1:2^{B_{p}}]$. Let $\mathbf{x}_{c,k}\in \mathbb{C}^{\tilde{L} \times 1}$ denote the transmitted communication signal, i.e., signal $\mathbf{x}_{c,k}$ contains $J$ randomly selected $L_p$-length pilot codeword. Thus, the signal component of CUs at given transmission chunks can be written as $\mathbf{Y}_c = \sum_{k=1}^{\bar{K}_c}\mathbf{x}_{c,k}\mathbf{h}_{c,k}$ where scalar $\bar{K}_c$ denotes the number of CUs at given chunk. Both common codebooks are generated from sub-sampled Discrete Fourier Transform (DFT) matrix and scaled by the power constraint.
\section{SSCS-UNISAC Decoder Design}\label{sec.3}
After encoding procedures of CUs and SUs, the overlapped signals from all users at given chunk can be written as:
\begin{equation}
	\label{eq:4}
	\begin{aligned}
		\mathbf{Y} &= \mathbf{Y}_c+\mathbf{Y}_s+\mathbf{N}, \\
		& =\sum_{k=1}^{\bar{K}_c}\mathbf{x}_{c,k}\mathbf{h}_{c,k}+\sum_{k=1}^{\bar{K}_s}\mathbf{x}_{s,k}\mathbf{h}_{s,k}+\mathbf{N},
	\end{aligned}
\end{equation}
where matrix $\mathbf{N}$ is the AWGN following $\mathcal{CN}(\mathbf{0},\sigma^2\mathbf{I})$. All chunks share identical decoding procedures due to their independence. Reminding that the CUs and SUs are divided not only in codeword domain i.e., $\mathbf{A}_s, \mathbf{A}_c$ but also in power domain, i.e., power ratio $\beta$. In our case, the pilot codeword length of SUs is much longer than that of CUs offering more spreading gain. Therefore, the proposed decoder can work in a manner of successive interference cancellation (SIC) with treating interference as noise (TIN). 

\subsection{Iterative Decoding on CUs' Signal Component}
When the receiver decodes the signal component of CUs, the AWGN and the signals of SUs are treated as pseudo noise $\mathbf{N}_{c,p}$ with variance of $\sigma^2_{c,p} \approx \sigma^2+\frac{\beta P \bar{K}_s}{|\mathcal{A}_s|\tilde{L}}$:
\begin{equation}
	\label{eq:5}
	\begin{aligned}
	\mathbf{Y}&=\underbrace{\sum_{k=1}^{\bar{K}_c}\mathbf{x}_{c,k}\mathbf{h}_{c,k}}_{\text{CUs' Signal Component}}+\underbrace{\sum_{k=1}^{\bar{K}_s}\mathbf{x}_{s,k}\mathbf{h}_{s,k}+\mathbf{N}}_{\mathbf{N}_{c,p}, \sigma_{c,p}^2\approx \sigma^2+\frac{\beta P \bar{K}_s}{|\mathcal{A}_s|\tilde{L}}}.
	\end{aligned}
\end{equation}
\subsubsection{Activity Detection and Channel Estimation (ADCE)}
For signal model $\mathbf{Y}=\sum_{k=1}^{\bar{K}_c}\mathbf{x}_{c,k}\mathbf{h}_{c,k}+\mathbf{N}_{c,p}$, let $\mathbf{Y}(j)=\mathbf{Y} \left( (j-1)L_p+1:jL_p,:\right), j\in[1:J]$ denote the received signal at the $j$-th chunk transmission. Following the spirit of simultaneous orthogonal matching \cite{ODMA}, the receiver conducts compressive sensing (CS) decoding following procedures to achieve ADCE at all chunks:
\begin{itemize}
	\item[-]Initialization: $\mathbf{R}\leftarrow \mathbf{Y}(j)$, $\mathcal{A}\leftarrow \emptyset$
	\item[-]Activity Detection (AD): 
	
	$i\leftarrow \mathop{\arg\max}\limits_{i\in\{1,2,\ldots,2^{B_p}\}}\frac{\|\mathbf{R}^{\mathrm{H}}\mathbf{A}_{c}(:,i_k)\|_2}{\|\mathbf{A}_{c}(:,i_k)\|_2}$, $\mathcal{A} \leftarrow \mathcal{A}\cup i$;
	\item[-]Residual Update (RU): 
	
	$\mathbf{\Phi}\leftarrow \mathbf{A}_{c}\left(:,\{\mathcal{A}\}\right)$,
	$\mathbf{R}\leftarrow \left(\mathbf{R}-\mathbf{\Phi}\mathbf{\Phi}^{\dagger}\mathbf{Y}(j)\right )$
	\item[-]Repeat the AD and RU procedures to obtain the final $\mathcal{A}$
	\item[-]Channel Estimation (CE):
	
	 $\{\tilde{\mathbf{h}}_{c,k}\},k\in [1:\bar{K}_c]\leftarrow \mathbf{A}_{c}\left(:,\{\mathcal{A}\}\right)^{\dagger}\mathbf{Y}(j)$.
\end{itemize}
Moore-Penrose inverse is denoted by $\left(\cdot\right)^{\dagger}$ which dominates the computational complexity of ADCE around $\mathcal{O}(ML_p2^{B_p})$.

Since we assume the time of transmission $L$ is within the channel coherence time, the channel coefficient remains unchanged during each chunk of transmission. Meanwhile, there will be ambiguity issue of randomized order in the AD results among chunks, which will be solved by the later segment stitching procedure.
\begin{figure}[htp]
	\centering
	\includegraphics[width=3.5in]{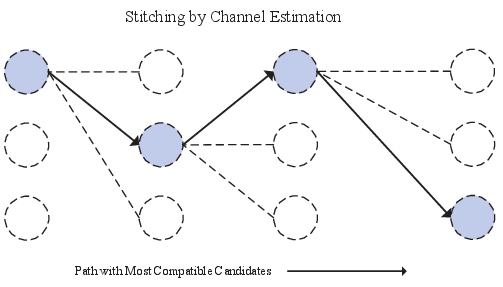}
	\caption{Illustration of stitching segments by channel estimation: Each selected candidate at next node is determined by evaluating the resemblance of channel estimation via \eqref{eq:6}, i.e., The channel coefficients can be treated as natural user tags due to the random nature of the channel conditions.}
	\label{MIMO_Stitching}
\end{figure}
\subsubsection{Segment Stitching by Channel Estimation}
After ADCE, the receiver obtains the information of all active pilot codewords and corresponding channel estimation. However, since the restored AD results have randomized order, the receiver needs to filter out the AD results of single CU to form a valid transmitted signals and restore the information. 

To fulfill this goal, the receiver treats the channel vectors as user tags of CUs and stitches a set of potential sub-frames together by evaluating the resemblance of the CE results across chunks:
\begin{equation}
	\label{eq:6}
	\begin{aligned}
		\arg \min_{\bar{\mathbf{h}}_{c,k,j}} \left \| \bar{\mathbf{h}}_{c,k,i} -\bar{\mathbf{h}}_{c,k,j}  \right \|_2^2,
	\end{aligned}
\end{equation}
where the normalized $\bar{\mathbf{h}}_{c,k,i}$ denotes the current node waiting for a match and the normalized $\bar{\mathbf{h}}_{c,k,j}$ denotes the node to be matched at the next node. Each by each via \eqref{eq:6}, the receiver filters out a list of $\bar{K}_c$ AD result frames. We illustrate the stitching procedure in Fig. \ref{MIMO_Stitching}. The complexity of searching and stitching scales around $\mathcal{O}(\bar{k}_cJ^2)$.
\subsubsection{Single Input Soft Output (SISO) Decoder}
For channel code decoding, the receiver starts by estimating likelihood information of each bits and then calculate the corresponding log-likelihood ratio (LLR). To calculate the LLR of the bits in a segment of singe CU form the observations at $j$-th sub-frame, we first form the noisy observations from the $k$-th CU by subtracting the signal from other CUs at given chunk:
\begin{equation}
	\mathbf{Y}_{j,k} = \mathbf{Y}_{j} - \sum_{k'\neq k} \mathbf{x}_{k'}\tilde{\mathbf{h}}_{k'} \approx \mathbf{x}_{k}\tilde{\mathbf{h}}_{k} +\mathbf{N}_j.
\end{equation}
Then, the likelihood of different pilot codewords of $k$-th CU can be estimated as:
\begin{equation}
	\begin{aligned}
		P(\mathbf{x}_k &=\mathbf{A}_c(:,i)\mid\mathbf{Y}_{j,k}), i=1,\ldots,2^{B_p} \\ &\propto\exp\left(\underbrace{-\frac{\|\mathbf{Y}_{j,k}-\mathbf{A}_c(:,i)\tilde{\mathbf{h}}_{c,k}\|_\mathcal{F}^2}{\sigma_{c,p}^2}}_{\beta_{k,j,i}}\right),
	\end{aligned}
\end{equation}
where $\|\cdot\|^2_{\mathcal{F}}$ denotes the Frobenius norm. Let $\mathbf{b}_{c,k}(n)$ denote the $n$-th bit in the segment and $\mathrm{bin}(i,n)$ denote the $n$-th bit of the binary expression of integer $i$. Thereupon, the LLR of the $n$-th bit in the binary segment vector can be calculated by:
\begin{equation}
	\begin{aligned}
		\mathrm{LLR}_n& \triangleq\log\frac{P\big(\mathbf{b}_{c,k}(n)=0\mid\mathbf{Y}_{j,k}\big)}{P\big(\mathbf{b}_{c,k}(n)=1\mid\mathbf{Y}_{j,k}\big)} \\
		&=\log\frac{\sum_{i:\mathrm{bin}(i,n)=0}\exp{(\beta_{k,j,i})}}{\sum_{i:\mathrm{bin}(i,n)=1}\exp{(\beta_{k,j,i})}} \\
		&\approx\max_{i:\mathrm{bin}(i,n)=0}\beta_{k,j,i}-\max_{i:\mathrm{bin}(i,n)=1}\beta_{k,j,i}.
	\end{aligned}
\end{equation}
Bit by bit, one can obtain all the LLR information of the $B_p$-bit. The SISO procedure is the process of non-linear demodulation whose complexity scales as $\mathcal{O}(2^{B_p})$ for single codeword and $\mathcal{O}(J\bar{K}_c2^{B_p})$ in total. Subsequently, the receiver inputs the estimated LLRs into the channel code decoder and then conduct parity check to verify whether the channel code is correctly decoded.
\subsubsection{Successive Interference Cancellation (SIC)} 
The receiver treats the parity check passed codeword as the correctly decoded messages and then re-encoded and re-modulate the parity check passed binary messages to conduct channel estimation refinement:
\begin{equation}
	\label{eq:7}
	\mathbf{H}_{\mathrm{refine}}=\left(\mathbf{X}_{\mathrm{pass}}^{\mathrm{H}}\mathbf{X}_{\mathrm{pass}}+\sigma_{c,p}^2\mathbf{I}_{K_{\mathrm{pass}}}\right)^{-1}\mathbf{X}_{\mathrm{pass}}^{\mathrm{H}}\mathbf{Y},
\end{equation}
where matrix $\mathbf{X}_{\mathrm{pass}}$ is the re-encoded and re-modulated $\tilde{L}$-length frame. Different to the channel estimator with $L_p$ length pilot in ADCE part, longer frame is used to estimate the channel, i.e., with $L=JL_p \gg L_p$, more precise channel estimation becomes possible. With refined channel estimation, the receiver subtracts $\mathbf{X}_{\mathrm{pass}}\mathbf{H}_{\mathrm{refine}}$ from the original noisy observation, i.e., $\mathbf{Y}\leftarrow \mathbf{Y}-\mathbf{X}_{\mathrm{pass}}\mathbf{H}_{\mathrm{refine}}$ and start the aforementioned decoding procedures iteration by iteration.
\subsection{Decoding on SUs' Signal Component}
Ideally, the signal component of CUs will be eliminated from the received noisy observations and only signal component of SUs and AWGN will exist. However, we consider a more generalized case where there may be $K_{c,md}$ missed detection of CUs in the noisy observation after SIC:
\begin{equation}
	\label{eq:11}
	\begin{aligned}
		\mathbf{Y}_{\mathrm{SIC}}&=\underbrace{\sum_{k=1}^{\bar{K}_s}\mathbf{x}_{s,k}\mathbf{h}_{s,k}}_{\text{SUs' Signal Component}}+\underbrace{\sum_{k=1}^{K_{c,md}}\mathbf{x}_{c,k}\mathbf{h}_{c,k}+\mathbf{N}}_{\mathbf{N}_{s,p},\sigma^2_{s,p}},
	\end{aligned}
\end{equation}
where signals of missed detection in communication and AWGN are treated as pseudo noise $\mathbf{N}_{s,p}$ with approximate variance of $\sigma^2_{s,p} \approx \sigma^2+\frac{(1-\beta) PK_{c,md}}{|\mathcal{A}_c|}$. To estimate the SUs' AOA, the receiver needs to conduct channel estimation and then detect the AOA information out of the estimation results. The solution to channel estimation from $\mathbf{Y}_{\mathrm{SIC}}$ is also a sparse recovery problem and can be tackled with by the referred CS decoder in Sec. III-A-1). Notably, the CS decoder utilizes the sensing common pilot codebook $\mathbf{A}_s$. After ADCE, the receiver estimates the AOA from estimation $\tilde{\mathbf{h}}_{s,k}$ in the spirit of multiple signal classification (MUSIC):
\begin{itemize}
	\item[-]Construct covariance matrix: $\mathbf{R}_s=\frac{1}{M}\tilde{\mathbf{h}}_{s,k}^{\mathrm{H}}\tilde{\mathbf{h}}_{s,k}$;
	\item[-]Eigen value decomposition: $\mathbf{R}_s=\mathbf{U}\mathbf{\Sigma}\mathbf{U}^{\mathrm{H}}$;
	\item[-]Formulate noise subspace with the eigen vectors of the $M-1$ smallest eigen values: 
	$\mathbf{R}_s=\mathbf{U}_s\mathbf{\Sigma}_s\mathbf{U}_s^{\mathrm{H}}+\mathbf{U}_N\mathbf{\Sigma}_N\mathbf{U}_N^{\mathrm{H}}$;
	\item[-]AOA search: $\tilde{\theta}_i \leftarrow \arg \max_{\theta}\frac{1}{\mathbf{a}^{\mathrm{H}}(\theta)\mathbf{U}_N\mathbf{U}_N^{\mathrm{H}}\mathbf{a}(\theta)}$, where $\mathbf{a}(\theta)$ is the array response of given angles in \eqref{eq:3}.
\end{itemize}
\section{Numerical Results}\label{sec.4}
\subsubsection{Parameter Setups And Benchmarks}
In this section, the numerical results of different power allocation and capacity under various system setups are illustrated. The benchmarks are achievable and optimistic (performance floor) bounds in work \cite{ISAC-URA}. The universal parameter setups are: For SUs, $B_s=15,B_{\mathrm{chunk}}=2,B_{s,p}=13,\tilde{L}=1250$; For CUs, $B=100,J_c=4,J=10,B_p=14,L_p=125,E=140$, low density parity check (LDPC) code is utilized as channel code. 
The LDPC decoding is done with standard belief propagation (BP) algorithm. The overall channel uses are fixed to $L=5000$ and the number of receiving antenna $M\in\{5,50,100\}$. Without loss of generality, the number of CUs and SUs are assumed to have $|\mathcal{A}_c|=|\mathcal{A}_s|$. The estimation targets are fixed to $\mathrm{PUPE}=0.1,\mathrm{AOAMSE}=5\times 10^{-4}$ in terms of the minimum-required energy-per-user. For the benchmarks, the performance capacity of the proposed scheme will be compared not only with the recently established bounds of UNISAC but also with various conventional protocols, including time-division multiple access (TDMA), TMDA-MUSIC, and ALOHA. For more detailed setups on conventional protocols, please refer to \cite{ISAC-URA}.
\begin{figure}[htp]
	\centering
	\includegraphics[width=3.5in]{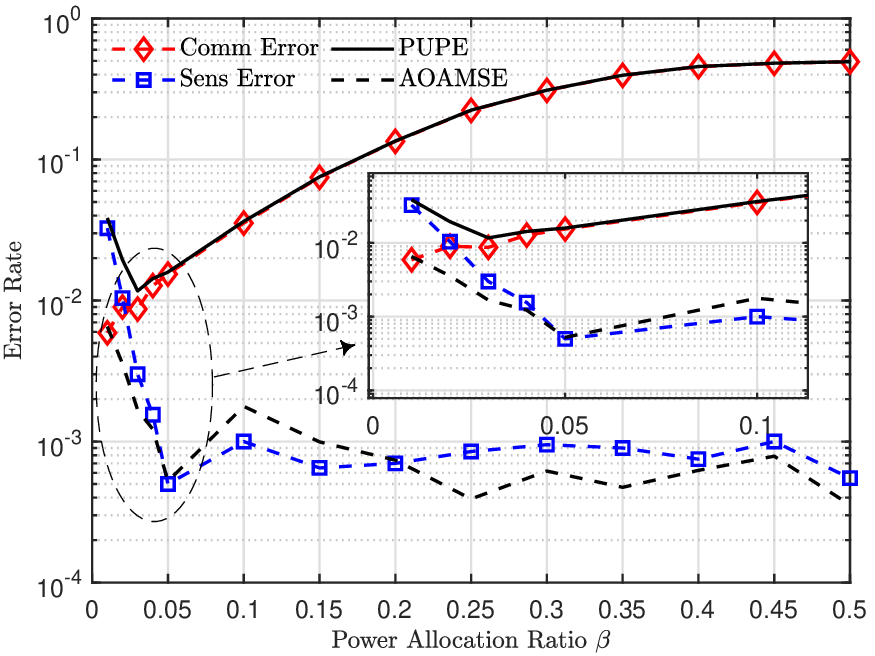}
	\caption{Illustration of error rate under different power allocation ratio $\beta$ with $E/N_0=20$dB, $|\mathcal{A}_c|+|\mathcal{A}_s|=200$, $M=5$.}
	\label{sim:power_allocation}
\end{figure}
\begin{figure}[htp]
	\centering
	\includegraphics[width=3.5in]{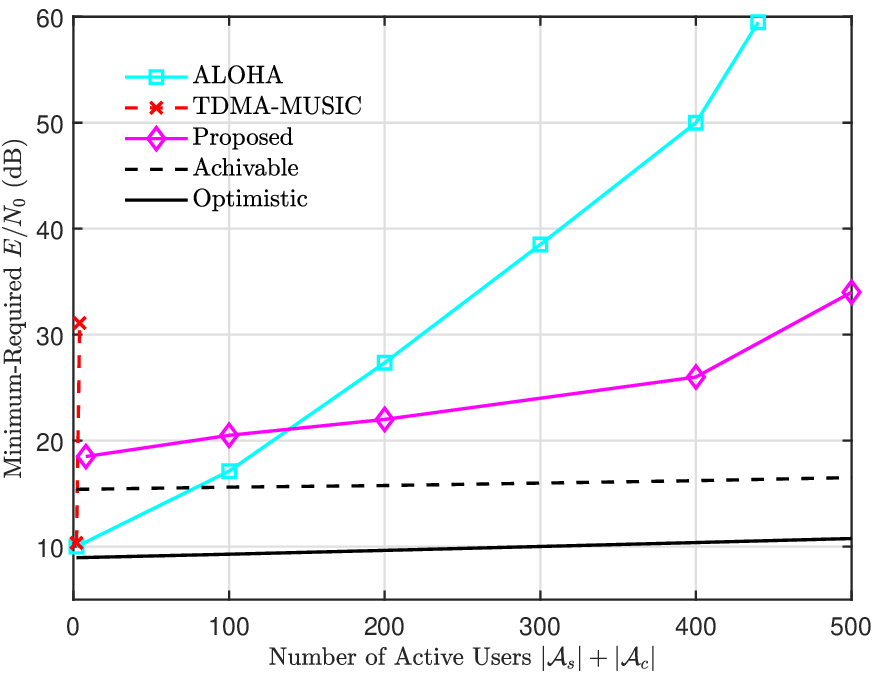}
	\caption{Minimum-required energy-per-user $E/N_0$ (dB) with a small number of receiving antennas, $M=5$. The benchmarks \cite{ISAC-URA} include the theoretically achievable and optimistic (performance floor) bounds of UNISAC, along with conventional protocols (ALOHA and TMDA-MUSIC).}
	\label{sim:small_antenna}
\end{figure}
%\begin{figure*}[htp]
%	\centering
%	\includegraphics[width=7in]{capacity.eps}
%	\caption{Illustration of minimum required energy-per-user to reach estimation targets of $\mathrm{PUPE}=0.1$ and $\mathrm{AOAMSE}=5\times 10^{-4}$ compared with achievable and optimistic Bounds in \cite{ISAC-URA} under different number of receiving antenna.}
%	\label{sim:capacity}
%\end{figure*}

\subsubsection{Power Ratio Selection $\beta$}The power allocation ratio between CUs and SUs is essentially needed and vital for the pseudo noise variance calculation. We conduct extensive simulations to search out a recommended ratio $\beta$ in Fig. \ref{sim:power_allocation} where energy-per-user equals to $E/N_0=20$dB, total number of active user is $|\mathcal{A}_c|+|\mathcal{A}_s|=200$, the number of receiving antenna is $M=5$. One can observe a lowest PUPE point at $\beta=0.03$ and a point with AOAMSE around $5\times 10^{-4}$ at $\beta=0.05$. Meanwhile, the performance of sensing error and AOAMSE features fluctuation which is normal due to different interference level from communication error. Considering the target goal of 0.1 PUPE has been reached, the final power allocation ratio is fixed to $\beta=0.05$ and the simulations in the sequel are conducted accordingly. Although $\beta=0.05$ may only be near-optimal, the numerical results in the following sections demonstrate the effectiveness of this power division ratio setups.
\begin{figure}[htp]
	\centering
	\includegraphics[width=3.6in]{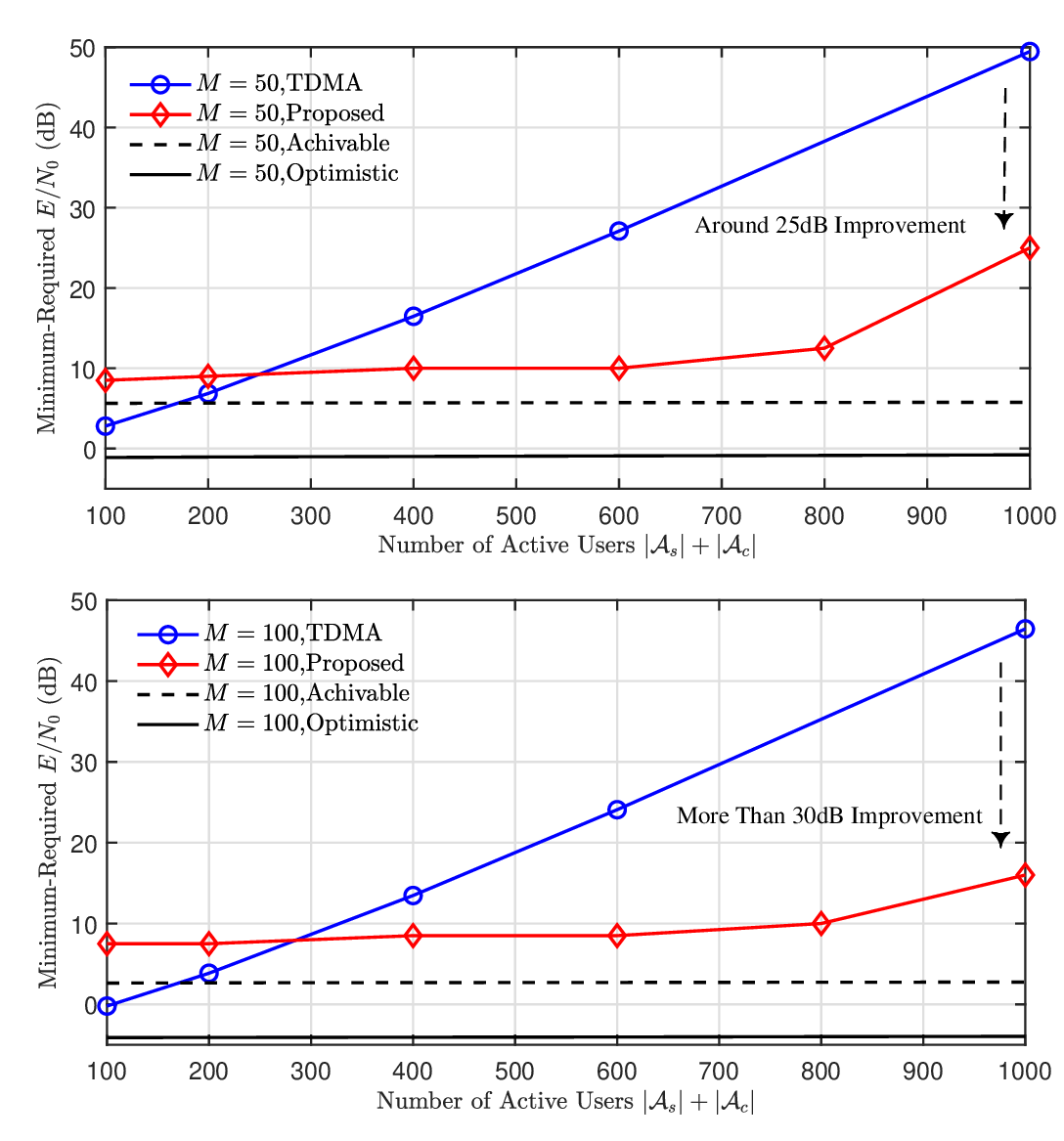}
	\caption{Minimum-required energy-per-user $E/N_0$ (dB) under large number of antennas, $M\in\{50,100\}$. The benchmarks \cite{ISAC-URA} include the theoretically achievable and optimistic (performance floor) bounds of UNISAC, as well as conventional protocol (TDMA).}
	\label{sim:large_antenna}
\end{figure}
\subsubsection{Capacity Performance With a Small Number of Antennas}
 In Fig.~\ref{sim:small_antenna}, the capacity performance of the proposed SSCS-UNISAC is illustrated, alongside theoretical benchmarks for both UNISAC and conventional protocols such as ALOHA and TMDA-MUSIC. Overall, the proposed scheme demonstrates greater robustness as the access density increases. Contrarily, TMDA-MUSIC fails to support the growing number of users, and the required minimum energy-per-user of ALOHA increases rapidly after $K_a=100$. The proposed scheme achieves a 30 dB capacity gain around 400 active users compared with ALOHA.
\subsubsection{Capacity Performance Under Large Number of Antennas} In Fig~\ref{sim:large_antenna}, We also compare the performance of the minimum required energy-per-user with a large number of antennas, where $M\in \{50,100\}$. The proposed scheme shows a significant performance gain compared to traditional protocols. For instance, there is a 25 dB performance improvement with 1000 active users and 50 receiving antennas, and more than a 30 dB performance gain when $M=100$. More importantly, by doubling the number of antennas from 50 to 100, the proposed scheme achieves nearly a 9 dB better performance, offering a favorable trade-off between resource utilization and energy consumption.
\section{Conclusion}\label{sec.5}
In this work, we propose a practical multiple access scheme called SSCS-UNISAC. The encoder and decoder designs for both CUs and SUs are presented, along with complexity analyses. The signal components of SUs and CUs are effectively separated in both the code and power domains, where the proposed non-linear decoder demonstrates excellent anti-interference and anti-noise capabilities. The SSCS-UNISAC scheme achieves favorable system capacity under both low and high numbers of receiving antennas, outperforming conventional schemes such as TDMA, ALOHA, and TDMA-MUSIC, particularly in high activity regions with a performance gain of 20-30 dB. However, the scheme's performance is influenced by the CS decoder design. Future work will focus on developing more powerful CS decoder designs and conducting further system analyses.
\section*{Acknowledgment}
This work is supported by NSFC projects (61971136,
61960206005), the Fundamental Research Funds for the
Central Universities (2242022k60001, 2242021R41149,
2242023K5003).
\balance

	%\newpage
	% your own custom macro containing the $\backslash${\tt{includegraphics}} command to make things
	%\bf{If you include a photo:}\vspace{-33pt}
	%\begin{IEEEbiography}[{\includegraphics[width=1in,height=1.25in,clip,keepaspectratio]{fig1}}]{Michael Shell}
	%Use $\backslash${\tt{begin\{IEEEbiography\}}} and then for the 1st argument use $\backslash${\tt{includegraphics}} to declare and link the author photo.
	%Use the author name as the 3rd argument followed by the biography text.
	%\end{IEEEbiography}

	\vfill
	
\end{document}